\documentclass[
reprint,
superscriptaddress,
 amsmath,
 amssymb,
 aps,
prb,
floatfix
]{revtex4-2}
\usepackage{graphicx}
\usepackage{dcolumn}
\usepackage{amsmath, amssymb}
\usepackage{dsfont}
\usepackage{comment}
\usepackage{bm}
\usepackage{placeins}
\usepackage{xstring}
\usepackage{cancel}
\usepackage{color}
\usepackage{afterpage}
\usepackage[caption=false]{subfig}

\usepackage{color}
\usepackage[normalem]{ulem}
\usepackage[colorlinks=true,citecolor=green,linkcolor=red,urlcolor=blue]{hyperref}
\usepackage{cleveref}
\usepackage{xcolor}
\usepackage[normalem]{ulem}
\usepackage{placeins}

\newcommand{\curl}[0]{\nabla \times}
\newcommand{\parent}[1]{\left( #1 \right)}
\newcommand{\rparent}[1]{\left[ #1 \right]}
\newcommand{\mc}[1]{\mathcal{{#1}}}
\newcommand{\scalar}[2]{\boldsymbol{#1}\cdot \boldsymbol{#2}}

\begin{document}

\preprint{APS/123-QED}

\title{Screened topological  plasmons in graphene plasmonic crystals}

\author{Andr\'e O. Soares}
\affiliation{Department and Center of Physics (CF-UM-UP), University of Minho, Campus of Gualtar, Braga, Portugal}
\affiliation{POLIMA---Center for Polariton-driven Light--Matter Interactions, University of Southern Denmark, Campusvej 55, DK-5230 Odense M, Denmark}

\author{Christos Tserkezis}
\affiliation{POLIMA---Center for Polariton-driven Light--Matter Interactions, University of Southern Denmark, Campusvej 55, DK-5230 Odense M, Denmark}
\affiliation{D-IAS---Danish Institute for Advanced Study, University of Southern Denmark, 5230 Odense M, Denmark}

\author{N. M. R. Peres}
\affiliation{Department and Center of Physics (CF-UM-UP), University of Minho, Campus of Gualtar, Braga, Portugal}
\affiliation{POLIMA---Center for Polariton-driven Light--Matter Interactions, University of Southern Denmark, Campusvej 55, DK-5230 Odense M, Denmark}
\affiliation{International Iberian Nanotechnology Laboratory (INL),  Avenida José Mestre, Braga, Portugal}

\begin{abstract}
We study topological effects in an one-dimensional 
plasmonic crystal formed by the screened plasmons emerging in a periodically modulated graphene sheet, placed on top of a metallic substrate. To this end, we develop the theory of quantization of screened plasmons, as appropriate for lossless graphene described by a Drude conductivity. By analyzing the resulting band structure, we show that the crystal sustains nontrivial topological bands, with quantized geometric phase. We further show that in a finite, open system, edge states appear within the band gap, which undergo a topological phase transition and merge with bulk states as the modulation increases. Our work provides a robust theoretical framework for the study of band structure and topology of layered media, and extends the possibilities for engineering two-dimensional materials with external modulation.
\end{abstract}

\maketitle

\section{\label{sec:Introduction} Introduction}

In metals, conduction electrons couple to incident light to generate collective charge density oscillations that confine electromagnetic
(EM) fields near the surface, forming excitations known as surface plasmons~\cite{polmanPlasmonicsOpticsNanoscale2005,perenboomElectronicPropertiesSmall1981}.
Plasmons are widely studied because of their extreme subwavelength field confinement and strong near-field enhancement, which enable nanoscale control of EM fields and light-matter interactions~\cite{gramotnevPlasmonicsDiffractionLimit2010,barnesSurfacePlasmonSubwavelength2003,polmanPlasmonicsApplied2008}. These excitations appear in a variety of setups, including  localized surface plasmons (LSPs) in metallic nanoparticles~\cite{liuRecentAdvancesPlasmonic2018,fanLightScatteringSurface2014}, or surface plasmon polaritons (SPPs) in extended metal-dielectric interfaces~\cite{bashevoyGenerationTravelingSurface2006}, and more recently in two-dimensional (2D) conducting surfaces~\cite{yoonPlasmonicsTwodimensionalConductors2014}.

In the latter, both the EM fields and the carriers are confined to two dimensions, substantially changing the properties of the collective behavior, allowing for plasmons with lower frequencies and longer propagation lengths~\cite{andressUltraSubwavelengthTwoDimensionalPlasmonic2012,sternPolarizabilityTwoDimensionalElectron1967}. 
Of high theoretical, experimental and technological interest are plasmons on a 2D graphene sheet, usually called graphene surface plasmons (GSPs). Propagating plasmons in graphene appear in the mid-infrared (IR) range~\cite{jablanPlasmonicsGrapheneInfrared2009} and showcase long propagation lengths with high level of field confinement~\cite{bonaccorsoGraphenePhotonicsOptoelectronics2010,woessnerHighlyConfinedLowloss2015,stauberPlasmonicsDiracSystems2014}. In addition, GSPs are highly tunable by different techniques, like gating~\cite{feiGatetuningGraphenePlasmons2012,chenOpticalNanoimagingGatetunable2012}, screening~\cite{alcaraziranzoProbingUltimatePlasmon2018,principiAcousticPlasmonsComposite2011}, doping~\cite{juGraphenePlasmonicsTunable2011} or other chemical means~\cite{takamuraPlasmonControlDriven2019,grigorenkoGraphenePlasmonics2012}, that can be explored in optoelectronic and sensing applications in the mid-IR range~\cite{rodrigoMidinfraredPlasmonicBiosensing2015,liGraphenePlasmonEnhanced2014,vasicLocalizedSurfacePlasmon2013}.

Compared to graphene deposited on a dielectric substrate, GSPs in the vicinity of a metal sheet can achieve much higher vertical confinement and momentum~\cite{alcaraziranzoProbingUltimatePlasmon2018,alonso-gonzalezAcousticTerahertzGraphene2017,goncalvesPlasmonicsLightMatter2020}.
In the former systems, the plasmon dispersion grows with the square root of the wave vector, but in the presence of a nearby metal the electric field of the plasmon excitations is screened, effectively softening the dispersion relation~\cite{tatianag.rappoportTopologicalGraphenePlasmons2021,principiAcousticPlasmonsComposite2011,goncalvesIntroductionGraphenePlasmonics2016,rodriguezecharriQuantumEffectsAcoustic2019} which becomes linear with the plasmon momentum. These excitations are known as acoustic  GSPs, reminiscent of acoustic phonons in solids. It is also possible to excite such screened plasmons using two layers of graphene separated by some small distance \cite{ferreiraQuantizationGraphenePlasmons2020,goncalvesIntroductionGraphenePlasmonics2016}(which is not the same as bilayer graphene), although these setups also have an optical branch where the dispersion grows with the square root of the wave vector.

In analogy to photonics~\cite{joannopoulosPhotonicCrystalsMolding2011} and electronics, plasmonic systems can be designed and engineered in periodic patterns to realize  2D crystalline structures, called plasmonic crystals~\cite{goncalvesIntroductionGraphenePlasmonics2016}.  There are multiple routes to construct such patterned plasmonic systems with graphene, for instance by patterning graphene into nanoribbons~\cite{nikitinSurfacePlasmonEnhanced2012,zhaoStrongPlasmonicCoupling2015,straitConfinedPlasmonsGraphene2013,rappoportUnderstandingElectromagneticResponse2020} or by coupling it to metallic gratings~\cite{diasControllingSpoofPlasmons2017,bylinkinTightBindingTerahertzPlasmons2019,guoHybridGrapheneplasmonGratings2023,zhaoStrongPlasmonicCoupling2015}.
In particular, graphene can be periodically gated in order to modulate the carrier density $n$~\cite{zhaoStrongPlasmonicCoupling2015,xiongPhotonicCrystalGraphene2019,jinTerahertzPlasmonicsFerroelectricgated2013,baeumerFerroelectricallyDrivenSpatial2015}, or equivalently, the Fermi energy level. In such systems, the resulting periodic structure can couple plasmons with different momenta,
giving rise to plasmonic bands~\cite{juGraphenePlasmonicsTunable2011,mirandaTopologyOnedimensionalPlasmonic2024}. In recent years, several efforts have focused on
understanding how concepts of topology and band theory are manifested in diverse nanostructured
systems like photonic~\cite{mortensenTopologicalNanophotonics2019,lanBriefReviewTopological2022,wangShortReviewAllDielectric2022,haldanePossibleRealizationDirectional2008,kimTopologicalEdgeCorner2020},
magnonic~\cite{shindouTopologicalChiralMagnonic2013,mcclartyTopologicalMagnonsReview2022}, or plasmonic
crystals~\cite{cuerdaObservationQuantumMetric2024,cuerdaPseudospinorbitCouplingNonHermitian2024,dipietroObservationDiracPlasmons2013} , among others. Although pristine graphene is not topological, it has been shown that GSPs in patterned and gated graphene-based one-dimensional (1D) and 2D plasmonic crystals can display topological
bands~\cite{breyQuantumBandStructure2025,jinInfraredTopologicalPlasmons2017}.

Topological properties of plasmons can be well described using a semi--classical picture for the light--matter interactions taking
place~\cite{wangBandTopologyClassical2019,tatianag.rappoportTopologicalGraphenePlasmons2021,mirandaTopologyOnedimensionalPlasmonic2024,nikitinEdgeWaveguideTerahertz2011}, when the number of photons in the EM field is large or the light source driving the system is
coherent~\cite{ferreiraQuantizationGraphenePlasmons2020}. However, quantization of the EM field is necessary to simultaneously understand the plasmon properties when the EM field involves only a small number of photons, and purely quantum effects start to take place, for instance in modeling coupling to quantum emitters or studying quantum 
interference~\cite{fakonasPathEntanglementSurface2015,heeresQuantumInterferencePlasmonic2013,xuQuantumPlasmonicsNew2018,tameQuantumPlasmonics2013,manriqueQuantumPlasmonEngineering2017}. The high degree of field confinement and long lifetime of GSPs make them ideal candidates to explore quantum plasmonic effects, with possible applications to generation of entangled
plasmons~\cite{breyQuantumPlasmonsDouble2024,sunGrapheneSourceEntangled2022} and plasmon-based quantum
computing~\cite{alonsocalafellQuantumComputingGraphene2019,calajoNonlinearQuantumLogic2023,gonzalez-tudelaEntanglementTwoQubits2011}.

In this work, we explore how the topological properties of screened plasmons in graphene can be understood when these excitations are quantized. To this end, we build a quantum-mechanical framework to study quantized GSPs screened by a metal near the graphene sheet that hosts the plasmon excitations. The plasmon excitations in these systems can be studied using near-field scanning optical microscope techniques \cite{alonso-gonzalezAcousticTerahertzGraphene2017,lundebergTuningQuantumNonlocal2017} Based on similar approaches~\cite{ferreiraQuantizationGraphenePlasmons2020,raymerQuantumTheoryLight2020}, we obtain fully analytical expressions for the mode functions, normalization length and dispersion relation, dealing explicitly with the case of \textit{lossless modes}, which can be well approximated for GSPs at low temperatures and in highly doped graphene.
We detail how a modulation of the Fermi energy in the graphene sheet affects the resulting quantum Hamiltonian, explicitly deriving an interaction term that arises from the periodicity of the modulation, similar to the one reported recently in Ref.~\cite{breyQuantumBandStructure2025}. These results are applied to the case of a step-like modulation of the Fermi energy level of graphene, analogous to a Kronig-Penney 
potential~\cite{kronigQuantumMechanicsElectrons1931}, which can be achieved using a split gate. Finally, we analyze the topological properties of the plasmonic bands and how the modulation parameters influence the bulk topological invariant. Using a supercell approach, we study the spectrum of a finite plasmonic crystal embedded in vacuum, and verify the bulk--boundary correspondence that relates a non-trivial bulk topological invariant with the appearance of mid gap states localized at the system's boundaries.

The article is structured as follows: in section \ref{sec:Quantization}, a detailed treatment on the canonical quantization of plasmon modes is carried out. It is first done for a generic plasmonic system, then applied to the case of screened graphene plasmons and finally to the complete graphene plasmonic crystal. In section \ref{sec:BandStructure}, a short review on the diagonalization of Bogoliubov-de Gennes (BdG) Hamiltonians is presented, followed by the numerical results of exact diagonalization of the screened GSP spectrum. Section \ref{sec:TopologicalAspects} begins with a presentation of methods to compute topological invariants of the plasmon bands through the Zak (or Berry) phases. There follows a discussion on the mechanism of band inversion, and the topological phase transition marked by a bulk topological invariant. The section ends with a demonstration of mid-gap edge states in a finite-size plasmonic-crystal slab at the interface with vacuum.

\section{\label{sec:Quantization} Quantization of plasmon excitations}

\subsection{\label{subsec:TotalEnergy} Energy density in dielectric media}
In our study, we are concerned with a non-magnetic structured medium composed by stacking a metal, a graphene sheet, and a homogeneous dielectric. To derive an expression for the total EM energy, we first consider the local energy density $u$. From Poynting's theorem~\cite{griffithsIntroductionElectrodynamics2023}, $u$ has both electric and magnetic contributions, which can be expressed as
\begin{equation}
    u = \int_{-\infty}^{t}dt'\,\boldsymbol{H}\cdot\partial_{t}\boldsymbol{B}+\boldsymbol{E}\cdot\partial_{t}\boldsymbol{D},
    \label{eq:EnergyDensity}
\end{equation}
where $\boldsymbol{E}$ and $\boldsymbol{D}$ represent the electric and displacement fields. We will consider non-magnetic media, such that the magnetic field $\boldsymbol{H}$ and induction $\boldsymbol{B}$ are simply related by $\boldsymbol{B}=\boldsymbol{H}/\mu_0$, where $\mu_{0}$ is the vacuum permeability.
In such systems, the electric permittivity $\epsilon_r$ has a position-$\boldsymbol{r}$ dependence, so we may express the displacement field $\boldsymbol{D}$ in frequency space
\begin{equation}
    \boldsymbol{D}(\boldsymbol{r},\omega)=\epsilon_{0}\epsilon_{r}(\boldsymbol{r},\omega)\boldsymbol{E}(\boldsymbol{r},\omega),
    \label{eq:ElectricDisplacement}
\end{equation}
where $\omega$ is an angular frequency and $\epsilon_0$ the vacuum permittivity.

In what follows we consider the Weyl gauge, with vanishing electrostatic potential $\phi = 0$. Thus, the fields can be expressed only in terms of the vector potential $\boldsymbol{A}$ as
\begin{align}
\boldsymbol{E}(\boldsymbol{r},t)	&=-\partial_{t}\boldsymbol{A}(\boldsymbol{r},t) \label{eq:WeylGauge_E}\\
\boldsymbol{B}(\boldsymbol{r},t)	&=\curl \boldsymbol{A}(\boldsymbol{r},t).
\label{eq:WeylGauge_B}
\end{align}
The vector potential can also be decomposed in terms of mode functions via a Fourier transform
\begin{equation}
    \boldsymbol{A}(\boldsymbol{r},t)=\sum_{p}\alpha_{p}\mathrm{e}^{-i\omega_p
t}\boldsymbol{A}_{p}(\boldsymbol{r})+c.c.,
    \label{eq:ModeDecompostion}
\end{equation}
where \textit{c.c.} stands for the complex conjugate of the preceding expression,
$\alpha_{p}$ are the mode amplitudes and the label $p=(\boldsymbol{q},\lambda)$ includes both momentum, $\boldsymbol{q}$, and polarization, $\lambda$, labels. The normal modes are encoded in the fields $\mathbf A_p(\mathbf r)$. To carry on a canonical quantization, however, one is interested in the total energy $W$, rather then the local energy density $u$. To compute it, we insert
Eq.~\eqref{eq:ModeDecompostion} into Eq.~\eqref{eq:EnergyDensity} and integrate over space. We find two terms, $W=W_+ + W_-$ (details of the calculation can be found in Appendix~\ref{appendix:QuantizationCalculations}),
\begin{equation}
    W_{\pm} =\frac{1}{2}\sum_{p,p'}\alpha_{p'}\alpha_{p}e^{-i(\omega_{p}\pm \omega_{p'})t}\int\frac{d\boldsymbol{r}}{V}\boldsymbol{A}_{p'}\cdot\mc{D}^{\pm}\cdot\boldsymbol{A}^{\pm}_{p}(\boldsymbol{r})+c.c,
    \label{eq:TotalEnergyFull}
\end{equation}
with $\boldsymbol{A}^{+}_p(\boldsymbol{r})=\boldsymbol{A}_p(\boldsymbol{r})$, $\boldsymbol{A}^{-}_p(\boldsymbol{r})=\boldsymbol{A}_p(\boldsymbol{r})^{\star}$, and $\mc{D}^\pm$ a coupling operator with the following form
\begin{align}
   \mc{D}^{\pm}= &\mp\omega_{p}\omega_{p'}\parent{\frac{\omega_{p}\epsilon_{r}(\boldsymbol{r},\omega_{p})\pm\omega_{p'}\epsilon_{r}(\boldsymbol{r},\omega_{p'})}{\omega_{p}\pm\omega_{p'}}} \nonumber \\
   &+\frac{1}{\mu_{0}}\curl \curl .
   \label{eq:CouplingOperator}
\end{align}

By imposing Maxwell's equations in matter, we are able to find a generalized eigenvalue equation for the mode functions $\boldsymbol{A}_p(\boldsymbol{r})$,
\begin{equation}
    \epsilon_{0}\omega_{p}^{2}\epsilon_{r}(\boldsymbol{r},\omega_{p})\boldsymbol{A}_{p}(\boldsymbol{r})=\frac{1}{\mu_{0}}\curl{\curl{\boldsymbol{A}_{p}}}(\boldsymbol{r}).
    \label{eq:NonLinearEigenvalueEq}
\end{equation}
Solutions to the above equation obey a set of orthogonality relations  that are detailed in Appendix~\ref{appendix:QuantizationCalculations}. These relations imply that the integrals in Eq.~\eqref{eq:TotalEnergyFull} vanish, except for degenerate terms in $W_-$  that satisfy $\omega_p-\omega_{p'}=0$. These are the only terms that have a non-zero contribution to the total energy, and can be simplified to
\begin{align}
    W&=\frac{1}{2}\sum_{p} \epsilon_{0}\omega_{p}^2\alpha_{p}\alpha_{p}^{\star}\int d\boldsymbol{r} \boldsymbol{A}_{p}(\boldsymbol{r}) \cdot \boldsymbol{A}_{p}^{\star}(\boldsymbol{r}) \times\nonumber\\& \times \left[\frac{\partial}{\partial\omega}\parent{\omega\epsilon_{r}(\boldsymbol{r},\omega)}+\epsilon_{r}(\boldsymbol{r},\omega_{p})
    \right]+c.c,
    \label{eq:TotalEnergySimplified}
\end{align}
where the derivatives with respect to $\omega$ are to be evaluated at $\omega_p$. This result is in agreement with previous
literature~\cite{ferreiraQuantizationGraphenePlasmons2020,archambaultQuantumTheorySpontaneous2010,hansonQuantumPlasmonicExcitation2015}.

\subsection{\label{subsec:ModeQuantization} Canonical Quantization}

The simplified form for the total energy in Eq. \eqref{eq:TotalEnergySimplified} resembles the energy of several simple harmonic oscillators. The canonical quantization is achieved by promoting the mode amplitudes $\alpha_q$ and $\alpha_q^\star$ to quantum-mechanical ladder operators:
\begin{align}
    \alpha_{p} &\rightarrow  \sqrt{\frac{\hbar}{2\epsilon_{0}\omega_{p}V_{\mathrm{eff}}}}a_{p} \nonumber\\
    \alpha^\star_{p} &\rightarrow  \sqrt{\frac{\hbar}{2\epsilon_{0}\omega_{p}V_{\mathrm{eff}}}}a^\dagger_{p}.
    \label{eq:CanonicalQuantizationStep}
\end{align}
The operators $a_q/a_q^\dagger$ are bosonic ladder operators that destroy/create a plasmon with quantum number $q$, and satisfy the following equal-time commutation relation:
\begin{equation}
    [a_{p},a_{p'}^\dagger] = \delta_{p,p'}.
    \label{eq:CommutationRelation}
\end{equation}
The quantity $V_{\mathrm{eff}}$ is an effective ($d$-dimensional) volume necessary to ensure that the Hamiltonian obtained by inserting Eqs.~\eqref{eq:CanonicalQuantizationStep} into the expression for the total energy, Eq.~\eqref{eq:TotalEnergySimplified}, coincides with the familiar Hamiltonian for a collection of harmonic oscillators (c.f. Appendix \ref{appendix:QuantizationCalculations} for further details):
\begin{equation}
    H = \frac{1}{2}\sum_p \hbar\omega_q \parent{a_{p}a^\dagger_{p} + h.c.},
    \label{eq:HarmonicOscilatorHamiltonian}
\end{equation}
where $\textit{h.c.}$ stands for Hermitian conjugate. In doing this matching, we find
\begin{equation}
    V_{\mathrm{eff}} = \int d\boldsymbol{r}\boldsymbol{A}_{p} (\boldsymbol{r})\left[\epsilon_{r}(\boldsymbol{r},\omega_{p})+\frac{\omega_{p}}{2}\frac{\partial}{\partial\omega}\parent{\epsilon_{r}(\boldsymbol{r},\omega)}\right]\boldsymbol{A}_{p}^{\star}(\boldsymbol{r}).
    \label{eq:EffectiveVolume}
\end{equation}

\subsection{\label{subsec:GrapheneSPP} Graphene Surface Plasmons of screened nature}

In this section, we adapt these results to a structured dielectric medium that contains a graphene sheet placed in close proximity to a perfect metallic mirror, as depicted in Fig ~\ref{fig:ExperimentalSetup}. We consider that the graphene sheet lies on the $z=0$ plane and, for simplicity, that both the surrounding medium and the graphene sheet are homogeneous along the $x,y$ directions. The metallic layer is considered to be a semi-infinite plane, such that we can safely neglect non-local effects in the metal on the plasmon dispersion~\cite{diasProbingNonlocalEffects2018, lundebergTuningQuantumNonlocal2017}.

The graphene sheet is modeled as a 2D infinitesimally thin, two-sided surface, characterized by an optical conductivity dominated by the Drude contribution
\begin{figure}[t!]
    \centering
    \includegraphics[width=0.52\textwidth]{Graphics/Diagram_axis_HQ.pdf}
    \caption{Schematic illustration of the system supporting screened graphene plasmons. The blue layers surrounding graphene (honeycomb lattice) represent a dielectric, for example hexagonal boron nitride. The dielectric below graphene has a thickness of $d$, and sits on top of a metal substrate. The metallic inverted pyramid represents the tip of a near-field scanning optical microscope, used to detect graphene plasmons, and the red arrow pointing to the sample the incoming terahertz light that excites the graphene plasmons.}
    \label{fig:ExperimentalSetup}
\end{figure}
\begin{equation}
    \sigma(\omega)=\frac{D_0}{\Gamma-i\omega},
    \label{eq:DrudeConductivity}
\end{equation}
with $\Gamma$ being the relaxation rate that accounts for optical losses and $D_0=\mathsf{e}^2E_{\mathrm{F}}/\pi\hbar^2$ the Drude weight in graphene,
where $\mathsf{e}$ is the elementary charge and $E_{\mathrm{F}}$ is the Fermi energy. This approximation works well at low temperatures and highly doped graphene. For a distance $d$ between the metal and the graphene sheet, and a surrounding dielectric characterized by $\epsilon_d$, the permittivity is 
\begin{equation}
    \epsilon_r^0(\boldsymbol{r},\omega)=
    \begin{cases}
         \epsilon_d, &-d\leq z < 0 \,\land \, z>0 \\
         i\frac{\sigma (\omega)}{\epsilon_0 \omega}\delta(z), & z=0 
    \end{cases}
, \label{eq:DielectricFunction}
\end{equation}
where $\delta(z)$ represents the Dirac delta function. Due to translational invariance along the $x$ and $y$ directions, the mode decomposition of the vector potential can be expressed as
\begin{equation}
    \boldsymbol{A}(\boldsymbol{r},t)=\frac{1}{\sqrt{S}}\sum_{\boldsymbol{q},\lambda}\alpha_{\boldsymbol{q},\lambda}(t)e^{i\scalar qx}\boldsymbol{A}_{\boldsymbol{q},\lambda}(z)+c.c,
    \label{eq:NewModeDecomposition}
\end{equation}
where $S$ is the area of the graphene sheet, $\boldsymbol{x}\in \mathbb{R}^2$ a position vector in the graphene plane, and we have expanded the label $p= (\boldsymbol{q},\lambda)$ to contain a momentum label $\boldsymbol{q}=(q_x,q_y)$ and polarization $\lambda$. The plasmon mode functions and dispersion relation can be obtained by looking for $p$-polarized solutions of the electric field of the generalized eigenvalue equation~\eqref{eq:NonLinearEigenvalueEq}, together with Eq.~\eqref{eq:DielectricFunction} and appropriate boundary conditions, that are well known in the 
literature~\cite{mirandaTopologyOnedimensionalPlasmonic2024}. Assuming a quasi-static approximation $c \rightarrow\infty$ for the plasmon dispersion, and denoting $q=|\boldsymbol{q}|$, we may write the dispersion relation for screened graphene plasmons as
\begin{equation}
    \frac{\epsilon_d}{q}\parent{1+\coth{\parent{qd}}}+i\frac{\sigma (\omega)}{\epsilon_0 \omega}=0.
    \label{eq:DispersionRelation}
\end{equation}

In the presence of losses, the frequency of the plasmons is, generally, complex $\omega = \omega'+i\omega''$. However, typical values for the relaxation rate in graphene are quite small, on the order $\Gamma\approx4\,\mathrm{meV}$, such that we may consider that their effect on the dispersion is quite small $\omega''\ll\omega'$. With these considerations, the dispersion in Eq.~\eqref{eq:DispersionRelation} can be solved as
\begin{subequations}
    \begin{equation}
         \omega'= \sqrt{\frac{D_0}{\epsilon_d\epsilon_0}\frac{q}{1+\coth{(qd)}}}
         \label{eq:RealDispersion}
    \end{equation}
    \begin{equation}
        \omega''=-\frac{\Gamma}{2}
    \label{eq:ImaginaryDispersion}
    \end{equation}
    \label{eq:AGSPDispersionRelation}
\end{subequations}

In the remainder of the discussion, we will neglect the contributions of losses, as they simply introduce a constant decay of the plasmon modes. For small graphene--metal separation $qd\ll1$, the dispersion relation is linear, $\omega_q\propto q$, which exponentially tends to a square--root regime $\omega_q \propto \sqrt{q}$ as this distance increases. The mode functions take the following form
\begin{equation}
\label{eq:AGSPModeFunctions}
    \mathbf{A}_{\mathbf{q}}(z)=\begin{cases}
\begin{aligned}
i&\sinh(q(z+d))\frac{\mathbf{q}}{q} \\
&+\cosh(q(z+d))\boldsymbol{\hat{z}}
\end{aligned}, &0>z>-d \\
\\
\sinh(qd)e^{-qz}\parent{i\frac{\mathbf{q}}{q}-\frac{z}{|z|}\hat{\mathbf{z}}}, &z>0.
\end{cases}
\end{equation}
Since the mode functions depend only on $z$, the effective volume in Eq.~\eqref{eq:EffectiveVolume} becomes a normalizing length $L_{\mathrm{eff},q}$ given by
\begin{equation}
    L_{\mathrm{eff},q}=\frac{2\epsilon_{d}}{q}\left[\sinh^{2}(qd)+\frac{1}{2}\sinh(2qd)\right].
    \label{eq:AGSPEffLength}
\end{equation}

\subsection{\label{subsec:AGSP} Introducing a periodic modulation}

We now consider the effect of adding a periodic modulation to the Drude weight that governs the optical conductivity of the graphene sheet, such that $D(\boldsymbol{x}+\boldsymbol{R})=D(\boldsymbol{x})$ for some lattice vectors $\{\boldsymbol{R}\}$. Such a modulation can be realized by introducing an equal modulation to the Fermi energy of graphene. We then have to express the Drude weight as a Fourier series
\begin{equation}
    D(\boldsymbol{x})=\sum_{\boldsymbol{G}}D_{\boldsymbol{G}}e^{i\scalar Gx}.
    \label{eq:DrudeModulation}
\end{equation}
Here, $D_{\boldsymbol{G}}$  are the Fourier coefficients of the modulation and $\{\boldsymbol{G}\}$ the reciprocal lattice vectors associated to the crystalline structure, defined by $\boldsymbol{R}\cdot \boldsymbol{G}=2\pi  m, m\in\mathbb{N}$.  In turn, this will alter the structure of the dielectric function, which now reads
\begin{align}
    \epsilon_{r}(\boldsymbol{r},\omega) &=\epsilon_{d}+\sum_{\boldsymbol{G}}\epsilon_r^{\boldsymbol{G}}(z,\omega)e^{i\scalar Gx} \nonumber \\
    & = \epsilon^0_{r}(\boldsymbol{r},\omega)-\frac{\delta(z)}{\epsilon_{0}\omega^{2}}\sum_{\boldsymbol{G} \neq 0}D_{\boldsymbol{G}}e^{i\scalar Gx},
    \label{eq:ModulatedDielectricFunction}
\end{align}
where $\epsilon^0_{r}(\boldsymbol{r},\omega)$ is the same function given in Eq.~\eqref{eq:DielectricFunction}, which contains the $\boldsymbol{G}=0$ term.

In general, obtaining analytical solutions to the generalized eigenvalue problem defined in Eq.~\eqref{eq:NonLinearEigenvalueEq} for a periodically modulated dielectric profile is not straightforward, and the canonical quantization used previously cannot be applied directly. To proceed, we treat the modulation as a perturbation, and use the mode functions and dispersion relations of the unmodulated system, i.e. the ones obtained by solving 
Eq.~\eqref{eq:NonLinearEigenvalueEq} with only $\epsilon^0_{r}(\boldsymbol{r},\omega)$. Within this perturbative framework, the orthogonality relations no longer diagonalize the energy expression in Eq.~\eqref{eq:TotalEnergyFull}, and cross terms appear that couple modes whose momenta differ by a reciprocal lattice vector $\boldsymbol{G}$. Upon promoting the mode amplitudes to quantum-mechanical operators, these terms manifest as interaction processes between plasmons of different momenta (details of the derivation can be found in Appendix~\ref{appendix:QuantizationCalculations}). For the equilibrium configuration described in Sec.~\ref{subsec:GrapheneSPP}, the total Hamiltonian naturally separates as $H=H_0+H_{\text{int}}$. $H_0$ is given by Eq.~\eqref{eq:HarmonicOscilatorHamiltonian}, with the screened plasmon dispersion in Eq.~\eqref{eq:RealDispersion}, and describes the unperturbed plasmon modes, while $H_{\text{int}}$ accounts for their coupling induced by the modulation. These terms can be written as

\begin{subequations}
\label{eq:FullHamiltonian}
\begin{equation}
    H_0 = \frac{\hbar}{2}\sum_{\boldsymbol{k},\boldsymbol{G}}\omega_{\boldsymbol{k}+\boldsymbol{G}} \parent{a_{\boldsymbol{k}+\boldsymbol{G}} a^\dagger_{\boldsymbol{k}+\boldsymbol{G}} + h.c.}
    \label{eq:AGSPFull_Equilibrium}
\end{equation}
\vspace{2em}
\begin{widetext}
\begin{equation}
    H_{\text{int}}(t)=\sum_{\boldsymbol{k},\boldsymbol{G},\boldsymbol{G'}\neq\boldsymbol{G}}g_{\boldsymbol{k},\boldsymbol{G},\boldsymbol{G'}}\left[e^{i(\omega_{\boldsymbol{k}+\boldsymbol{G}}-\omega_{\boldsymbol{k}+\boldsymbol{G'}})t}a_{\boldsymbol{k}+\boldsymbol{G}}a_{\boldsymbol{k}+\boldsymbol{G'}}^{\dagger}+e^{i(\omega_{\boldsymbol{k}+\boldsymbol{G}}+\omega_{-(\boldsymbol{k}+\boldsymbol{G'})})t}a_{\boldsymbol{k}+\boldsymbol{G}}a_{-(\boldsymbol{k}+\boldsymbol{G'})}\right]+h.c.,
    \label{eq:AGSPFullHamiltonian}
\end{equation}
\end{widetext}
\end{subequations}

where $g_{\boldsymbol{k},\boldsymbol{G},\boldsymbol{G'}}$ is a coupling function and the momentum $\boldsymbol{k}$ is restricted to the first Brillouin zone (BZ). For the system considered in the previous section, this coupling takes the form
\begin{equation}
    g_{\boldsymbol{k},\boldsymbol{G},\boldsymbol{G'}} =\frac{\hbar}{2}\frac{D_{\boldsymbol{G}-\boldsymbol{G'}}}{D_0}f(\boldsymbol{k}+\boldsymbol{G})f(\boldsymbol{k}+\boldsymbol{G'}),
    \label{eq:CouplingTerm}
\end{equation}
with 
\begin{equation}
    f(\boldsymbol{q})=\left[\frac{D_{0}}{\epsilon_{0}\epsilon_{d}}\frac{q}{\parent{1+\coth\left(qd\right)}}\right]^{1/4}\frac{\boldsymbol{q}}{q} .
    \label{eq:FormFactor}
\end{equation}
The first term in Eq.~\eqref{eq:AGSPFullHamiltonian} describes scattering processes between two plasmons that exchange momentum $\boldsymbol{G}-\boldsymbol{G'}$. There are also anomalous terms proportional to  $a_{\boldsymbol{k}+\boldsymbol{G}}a_{\boldsymbol{k}+\boldsymbol{G'}}$ (and their Hermitian conjugates), which do not conserve particle number, as they involve the simultaneous creation/annihilation of plasmon pairs, and oscilate at approximately twice the plasmon frequency. These are known as counter-rotating terms and are usually neglected under the rotating-wave approximation \cite{foxQuantumOpticsIntroduction2006}, but become significant in the strong-coupling regime, where the modulation strength is comparable to the plasmon frequency.

The resulting Hamiltonian is derived in a time-dependent interaction picture, but it is more convenient to work with a time-independent picture. This can be achieved by applying a unitary transformation
\begin{equation}
    H^{(S)}=e^{-iH_{0}t/\hbar}H(t)e^{iH_{0}t/\hbar}.
    \label{eq:SchrodingerPicture}
\end{equation}
The operator $\exp(iH_{0}t/\hbar)$ commutes with $H_0$ and eliminates the time-dependent exponentials in Eq.~\eqref{eq:AGSPFullHamiltonian}, something that can be verified by its commutation relations with the operators $a_q/a_q^\dagger$.

\section{\label{sec:BandStructure} Band structure of screened graphene surface plasmons }

\subsection{\label{subsec:BogoliubovTransformation} Bogoliubov Transformation}

The Hamiltonian expressed in Eq.~\eqref{eq:FullHamiltonian} is a bosonic Bogoliubov-de Gennes (BdG) Hamiltonian, with pairing terms (particle--particle channels) and anomalous terms (particle--hole channels) governed by the same coupling factor $g_{\boldsymbol{k},\boldsymbol{G},\boldsymbol{G'}}$. Defining a vector $\psi_{k}=\left[a_{\boldsymbol{k}},\,\dots\,a_{\boldsymbol{k}+\boldsymbol{G}}\right]$, we can rewrite the Hamiltonian as
\begin{align}
    H&=\sum_{k}   \begin{bmatrix}
        \psi_{\boldsymbol{k}} &
        \psi^\dagger_{-\boldsymbol{k}} 
    \end{bmatrix}
    \mathcal{H}_{k}    \begin{bmatrix}
        \psi^\dagger_{\boldsymbol{k}} \\
        \psi_{-\boldsymbol{k}} 
    \end{bmatrix}, \hspace{0.5em} \mc{H}_k=\parent{\begin{matrix}
    A & B\\
    B^{\dagger} & A^{T}
  \end{matrix} } \\
\label{eq:BogoliubovHamiltonian}
[A]&_{\boldsymbol{G},\boldsymbol{G}} =\omega_{\boldsymbol{k}+\boldsymbol{G}} \hspace{1.5em}[A]_{\boldsymbol{G},\boldsymbol{G'}}=[B]_{\boldsymbol{G},\boldsymbol{G'}} =g_{\boldsymbol{k},\boldsymbol{G},\boldsymbol{G'}}, \nonumber
\end{align}
and $[B]_{\boldsymbol{G},\boldsymbol{G}}=0$. The diagonal blocks $A/A^T$ connect the particle--particle and hole--hole spaces, while the off-diagonal blocks $B$ connect the particle--hole spaces. The diagonal elements of the  blocks $B$ vanish because the modulation only couples plasmon modes with $\boldsymbol{G}\neq \boldsymbol{G'}$, and the $\boldsymbol{G}-\boldsymbol{G'}=0$ contribution is already absorbed in the diagonal components of the blocks $A$.

The diagonalization process requires finding a new basis set of creation/destruction operators $\gamma_{\boldsymbol{k}}=\left[b_{\boldsymbol{k}},\,\dots\,b_{\boldsymbol{k}+\boldsymbol{G}}\right]$, and a transformation $T_{\boldsymbol{k}}[\psi_{\boldsymbol{k}}\,\,\psi^\dagger_{-\boldsymbol{k}}]=[\gamma_{\boldsymbol{k}}\,\,\gamma^\dagger_{-\boldsymbol{k}}]$ such that $H=\sum_{k}\gamma_{k}^{\dagger} D_{k}\gamma_{k}$ with $D_k$ being a diagonal matrix~\cite{shindouTopologicalChiralMagnonic2013,gorenTopologicalZakPhase2018}. However, imposing bosonic commutation relations on the new set of operators $b_{\boldsymbol{k}}/b^\dagger_{\boldsymbol{k}}$ forces the transformation matrix $T_{\boldsymbol{k}}$ to be para-unitary. Thus, one needs to diagonalize $\Gamma \mc{H}_{\boldsymbol{k}}$, where $\Gamma$ is a diagonal matrix that takes $\pm 1$ values
in the particle/hole space.

Due to the presence of the anomalous pairing terms, the transformed operators contain contributions from both 
the particle and the hole spaces,
\begin{equation}
b_{n,\boldsymbol{k}}=\sum_{\boldsymbol{G}}\parent{\alpha^n_{\boldsymbol{k}+\boldsymbol{G}}a_{\boldsymbol{k}+\boldsymbol{G}}+\beta^n_{\boldsymbol{k}+\boldsymbol{G}}a_{-(\boldsymbol{k}+\boldsymbol{G})}^{\dagger}},
    \label{eq:BogoliubovOperators}
\end{equation}
where $\{\alpha^n_{\boldsymbol{k}}\}$ and $\{\beta^n_{\boldsymbol{k}}\}$ are coefficients that describe the transformation $T_{\boldsymbol{k}}$, and $n$ is a band index. Enforcing bosonic commutation relations on the new set of creation/annihilation operators $b^\dagger_{n,\boldsymbol{k}}/b_{n,\boldsymbol{k}}$ yields on the coefficients the constraint
\begin{equation}
    \sum_{G} \parent{|\alpha^n_{\boldsymbol{k+G}}|^2-|\beta^n_{\boldsymbol{k+G}}|^2}=1.
    \label{eq:BosonicCommutation}
\end{equation}

The real-space wavefunction profile associated to the quasiparticle operators $b_{n,\boldsymbol{k}}$ can be obtained by projecting the field operator $\Psi(\boldsymbol{x})=\sum_{\boldsymbol{q}}a_{\boldsymbol{q}}e^{i\scalar{q}{x}}$ onto these modes. This leads to the following particle--hole Nambu spinor
\begin{equation}
    \psi_{n,\boldsymbol{k}}(\boldsymbol{x})=\sum_{\boldsymbol{G}}\parent{\begin{matrix}\alpha^n_{\boldsymbol{k}+\boldsymbol{G}}e^{i(\boldsymbol{k}+\boldsymbol{G})\cdot \boldsymbol{x}}\\
-\beta^n_{\boldsymbol{k}+\boldsymbol{G}}e^{i(\boldsymbol{k}+\boldsymbol{G})\cdot \boldsymbol{x}}
\end{matrix}}.
\label{eq:WavefunctionSpinor}
\end{equation}
The relative minus sign in the hole component follows from the para-unitary normalization of the Bogoliubov transformation. In this basis, the inner product of these spinors is endowed with a metric $\sigma_z=\text{diag}(1,-1)$ that is consistent with the bosonic commutation relations, such that
\begin{equation}
    \langle\psi_{n,\boldsymbol{k}}|\psi_{n,\boldsymbol{k}}\rangle=\int d\boldsymbol{x}\,\psi_{n,\boldsymbol{k}}^\dagger \cdot \sigma_z\cdot \psi_{n,\boldsymbol{k}}=1.
    \label{eq:ScalarMetric}
\end{equation}

\subsection{\label{subsec:PlasmonicBands} Plasmonic bands}

\begin{figure}[t!]
         \centering
         \includegraphics[clip, trim=0.0cm 2cm 0.0cm 2cm, width=0.4\textwidth]{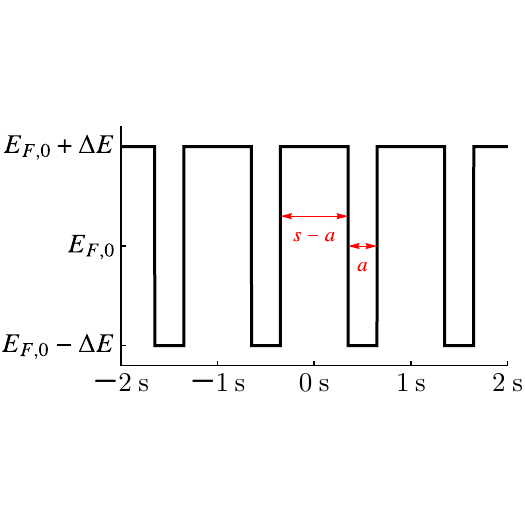}
    \caption{Modulation of the Fermi energy level in space, across four unit cells of length $s$. The modulation widths $a$ in the lower step of the Fermi level, and $s-a$ in the upper step, are marked with red arrows.}
    \label{fig:EFModulation}
\end{figure}

We will consider the case of a 1D modulation of the Fermi energy, aligned along the $\boldsymbol{\hat{x}}$ direction, and center the discussions around a Kronig-Penney modulation with a step-like structure of amplitude $2 \Delta E$, alternating between regions of high Drude weight $E_{\mathrm{F}}(x)=E_{F,0}+\Delta E_{\mathrm{F}}$, and lower regions $E_{\mathrm{F}}(x)=E_{F,0}-\Delta E_{\mathrm{F}}$, as shown in Fig.~\ref{fig:EFModulation}. As alluded to before, because the Drude weight is proportional to the Fermi energy in graphene, this is equivalent to setting a step modulation of the Drude weight as in Eq. \eqref{eq:DrudeModulation}.
The period of the modulation defines a unit cell of length $s$, consisting of a region of width $a$ in the lower step and a region of width $s-a$ in the upper step. Because this is a 1D modulation, the reciprocal lattice vectors can be expressed as $\boldsymbol{G}_n=2\pi n\boldsymbol{\hat{x}} /s$, $n\in \mathbb{Z}$ , and we will consider only plasmons with vanishing transverse momentum $k_y=0$.

To employ the Bogoliubov diagonalization, we first truncate the sum over reciprocal lattice vectors and numerically obtain the spectrum. Considering $N$ vectors, the Hamiltonian becomes a $2N \times 2N$ matrix, with the first $N$ eigenvalues $E_n$ corresponding to particle excitation energies, while the other eigenvalues, $-E_n$, correspond to hole excitations. We discard the last set of eigenvalues and focus on the positive energy part of the spectrum.

\begin{figure}[t!]
\centering
         \includegraphics[width=0.85\linewidth]{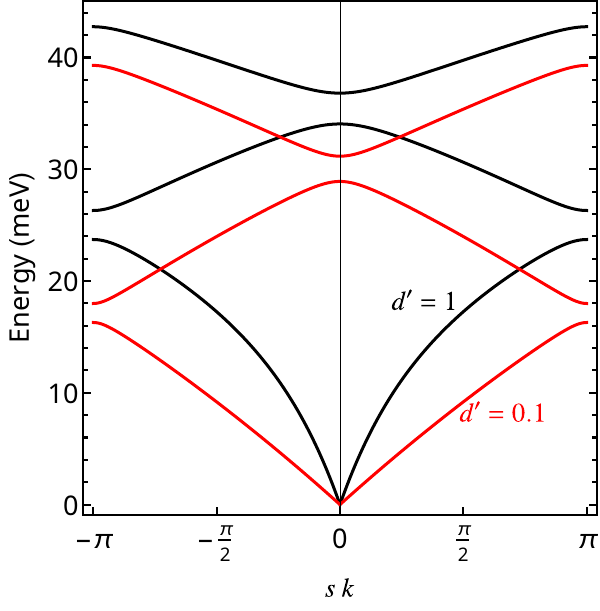}
         \caption{Spectrum of the Hamiltonian given in Eqs.~\eqref{eq:FullHamiltonian} for two values of graphene--metal separation $d^\prime=d/s$, with parameters $s=500$ nm, $a=250$ nm, $E_{\mathrm{F},0}=500$ meV and $\Delta E=E_{\mathrm{F},0}/5$. In black line, a large separation $d^\prime=1$ yields an approximately parabolic dispersion, and in red line a very small separation, $d^\prime=0.1$, produces a linear dispersion.}
         \label{fig:Spectrum}
\end{figure}

We show in Fig.~\ref{fig:Spectrum} the spectrum for the plasmonic crystal in two cases where we vary the graphene--metal separation $d$. Since the EM fields must vanish at the metal interface, the plasmon modes become confined in tighter spaces as $d$ becomes smaller, i.e. the screening effect becomes larger, leading to acoustic--like plasmons with high momentum \cite{diasProbingNonlocalEffects2018,principiAcousticPlasmonsComposite2011}.  It is possible then to observe a highly screened limit $d/s\ll1$, which shows a linear dispersion at low energies ~\cite{tatianag.rappoportTopologicalGraphenePlasmons2021,goncalvesIntroductionGraphenePlasmonics2016}, and a regime $d/s\approx1$ where the dispersion is approximately a square-root function. Indeed, the limit $d\rightarrow \infty$ agrees with the results for a system where graphene is placed between two infinite dielectrics. In addition, the bare dispersion in Eq. \eqref{eq:RealDispersion} grows with the separation $d$, meaning that at a fixed energy, screened plasmons have higher momentum than unscreened plasmons.

\section{\label{sec:TopologicalAspects} Topological Aspects of the Plasmonic Crystal}

In this section, we study the band topology of the plasmonic crystal outlined in the previous sections. Similar to electronic systems, the topological content of the spectrum of acoustic and photonic systems contains information about various properties. The central concept underlying the characterization of topological properties in these systems is that of Berry phase, and related quantities like the Berry connection and
curvature~\cite{vanderbiltBerryPhasesElectronic2018,haldanePossibleRealizationDirectional2008,khanikaevPhotonicTopologicalInsulators2013,zakBerrysPhaseEnergy1989}. In 1D systems, this phase is commonly referred to as the Zak phase, whose seminal work showed that this quantity can only take values of $0$ or $\pi$ in systems that have inversion symmetry~\cite{zakBerrysPhaseEnergy1989}.

\subsection{\label{subsec:InversionSymmetricZakPhase}Zak Phase in Inversion-Symmetric Systems}

The Zak phase for a 1D crystal system, where the crystal momentum evolves adiabatically over a band of index $n$, is defined in the following gauge-invariant form
\begin{equation}
    \gamma_n = -i\int_{-\pi/s}^{\pi/s} \langle \mathbf{u}_{n,k}(x)|\partial_k|\mathbf{u}_{n,k}(x) \rangle dk\,\,,
    \label{eq:ZakPhase_Def}
\end{equation}
where $\mathbf{u}_{n,k}(x)$ is the periodic Bloch component of the wavefunction, which satisfies $\mathbf{u}_{n,k}(x+s)=\mathbf{u}_{n,k}(x)$ and $ \psi_{n,k}(x)=e^{i k x}\boldsymbol{u}_{n,k}$. Like the plasmon wavefunctions defined in the previous section (see Eq.~\eqref{eq:WavefunctionSpinor}), the scalar products in Eq.~\eqref{eq:ZakPhase_Def} are also weighed by a $\sigma_z$ metric, as shown in Eq.~\eqref{eq:ScalarMetric}.

Although the Zak phase can generally take on any values, in inversion-symmetric systems it is quantized to take on values of $0$ or $\pi$. This value is actually connected to the product of the inversion-operator eigenvalues at time-reversal invariant momenta $k_T$~\cite{hughesInversionsymmetricTopologicalInsulators2011,liuZakPhaseCalculation2021,marti-sabateZaksPhaseNonSymmetric2021}. In one dimension, there are two such momenta $k_T=0$ and $k_T=\pi/s$, so we can use the following expression
\begin{equation}
    e^{i\gamma_n}=\xi_n (0)\xi_n (\pi/s)\,\,,
    \label{eq:ZakPhase_Parity}
\end{equation}
where $\xi_n (k_T)=\langle \psi_{n,k_T} (\boldsymbol{x}) | \mc{I}| \psi_{n,k_T} (\boldsymbol{x})\rangle = \pm1$ is the eigenvalue of the inversion operator, i.e. the parity of the wavefunction, that acts as
\begin{equation}
    \mc I \,\boldsymbol{\psi}_{n,k}(x)=\boldsymbol{\psi}_{n,k}(-x).
    \label{eq:ParityOperator}
\end{equation}
Equation~\eqref{eq:ZakPhase_Parity} allows us to compute the Zak phase without the integration over the reciprocal unit cell that appears in Eq.~\eqref{eq:ZakPhase_Def}.

\subsection{\label{subsec:Z2TopologicalInvariant} Topological Invariant and Parity Swapping}

In the previous section, the importance of the parity eigenvalue at the band center $(k=0)$ and edge ($k=\pi/s$) was outlined, since it is restricted by inversion symmetry to take values $\xi=\pm1$.
In electronic systems, the topological nature of $\gamma_n$ of each band means that this quantity is protected against small deformations of the Hamiltonian, for instance against disorder in the system or changes to the modulation profile. However, if the deformation is strong enough to close the gap between two adjacent bands, then the Berry connection becomes undefined and the system can undergo a topological phase transition. When this happens, the Zak phase of each band involved in the crossing can change. In inversion-symmetric systems, this means that the bands that cross can interchange their parity eigenvalue.

\begin{figure}[t!]
\begin{minipage}[b]{0.45 \textwidth}
    \includegraphics[width=\linewidth]{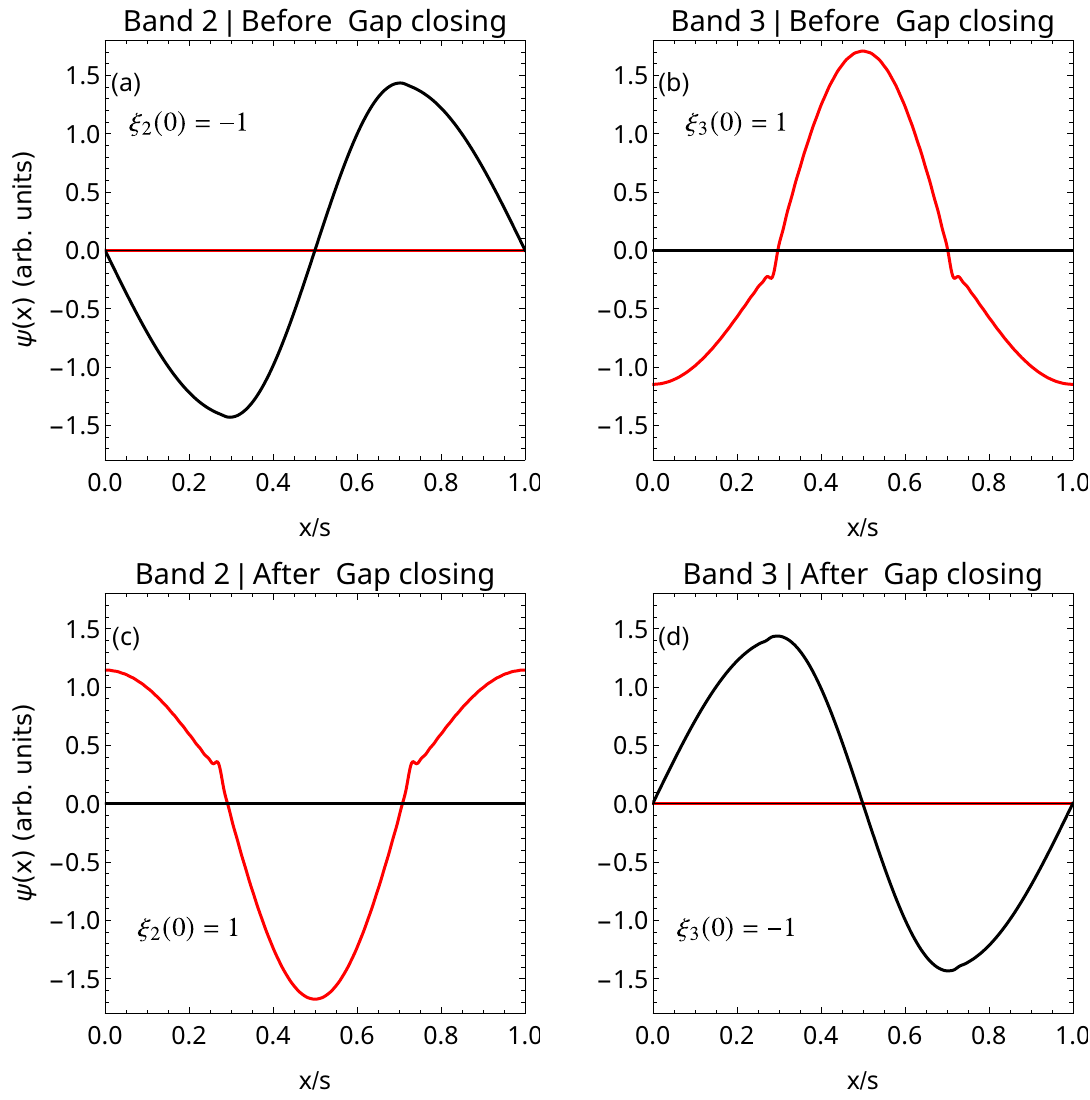}
    \end{minipage}
\hfill
\begin{minipage}[b]{0.45\textwidth}
\hfill
\centering
\includegraphics[width=\linewidth]{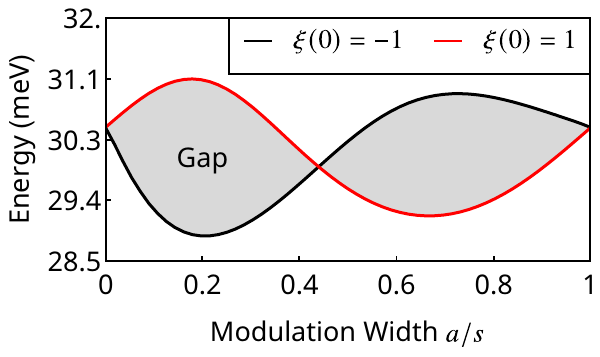}
\hfill
\end{minipage}
\hfill
    \caption{Top Panel: Profile of the particle-component (first entry in Eq.~(\ref{eq:WavefunctionSpinor})) of the wavefunction spinor (real part in red, imaginary part in black) before (plots (a) and (b)) and after (plots (c) and (d)) a topological phase transition. The parities $\xi_n$, $n=2,3$, of the second and third bands are denoted in each panel, and are swapped in the two phases. Bottom Panel: Dispersion of the states from bands 2 and 3 at the BZ center $k=0$, with a red (black) color representing the even (odd) parity of the wavefunctions, and shaded area highlighting the gap between the bands. These two bands become gapless at $k=0$ as the modulation width increases past a critical point $a_c \approx  0.44 \,s$ at fixed $\Delta E = E_{\mathrm{F},0}/5$.}
\label{fig:ParitySwapping}
\end{figure}

These considerations can be transposed to systems with bosonic quasiparticles of diverse
natures~\cite{xiaoGeometricPhaseBand2015,shindouTopologicalChiralMagnonic2013,gorenTopologicalZakPhase2018}.
Here, we observe this effect for the screened plasmonic crystal, by studying how the band topology changes as the modulation profile is tuned. It is necessary that the deformation of the Hamiltonian respects inversion symmetry, in order to maintain a quantized Zak phase, and that it can also induce band crossings, which can lead to a topological phase transition. It can be easily shown \cite{breyQuantumBandStructure2025} that both the modulation width $a$ and depth $\Delta E$ fulfill both of these requirements. However, this is not the case for the graphene-metal separation parameter $d$. Although tuning this parameter does deform the spectrum, as shown in figure \ref{fig:Spectrum}, it cannot induce any band crossings in the crystal. We focus on the relation between $\gamma_n$ with the modulation width $a$, and we exemplify by inspecting the second gap of the crystal.  This gap closes at a critical value of $a_{c}\approx 0.44\,s$, when the second and third bands cross at the band center.

In Fig.~\ref{fig:ParitySwapping} (top panel) we show the wavefunction profile for the second and third bands at the band center, when $a<a_c$ (before the second gap closes), and when $a>a_c$ (after the gap closes). When going from $(a)$ to $(c)$, or from $(b)$ to $(d)$, the parity
$\xi_n(0)$ changes sign. Since the bands do not touch at the band edge, then $\xi_n(\pi/s)$ remain unchanged for both bands. From Eq.~\eqref{eq:ZakPhase_Parity}, this means that $\gamma_2$ and $\gamma_3$ must also have their values swapped. In the bottom panel of Fig.~\ref{fig:ParitySwapping} this behavior is also depicted. We show the dispersion of the second and third states at the BZ center $k=0$, and color it according to the parity of the wavefunctions.

\begin{figure}[t!]
\centering
\begin{minipage}[b]{0.45 \textwidth}
\centering
  \includegraphics[clip, trim=1.0cm 3.3cm 0.5cm 3.3cm, width=\textwidth]{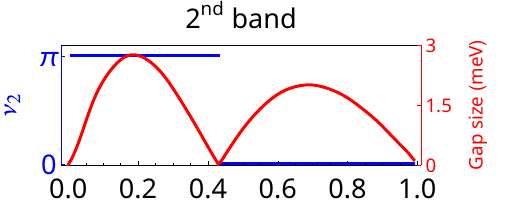}
\end{minipage}
\hfill
\begin{minipage}[b]{0.45\textwidth}
\centering
  \includegraphics[clip, trim=1.0cm 2.5cm 0.5cm 3cm, width=\textwidth]{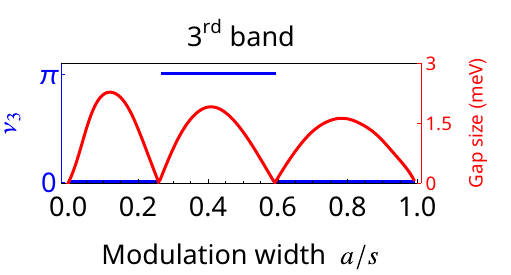}
\end{minipage}
\hfill
\caption{Phase diagram of the topological invariant $\nu$ (blue) as a function of the modulation width $a$. In red, the value of the gap just above the second (top) and third (bottom) gap. As the modulation width changes, the plasmon bands become gapless at some critical values of $a$, allowing for an exchange of Zak phase between the crossing bands. The second gap closes once.}
\label{fig:TopoInv}
\end{figure}

In analogy to electronic systems, we can build a $\mathbb{Z}_2$ topological invariant $\nu_n$ for each band, by summing over all the $\gamma_{n'}$ ($\text{mod} \,2\pi$), with $n'$ less than or equal to $n$ ~\cite{fuTopologicalInsulatorsInversion2007,breyQuantumBandStructure2025}. Whenever any gap below the band closes, the Zak phases of the bands that touch are exchanged, but leave the total sum unaltered. Equivalently, we can define
\begin{equation}
    e^{i \nu_n} = e^{i\sum_{n'\leq n} \gamma_{n'}}= \prod_{n'\leq n} \xi_{n'} (0)\xi_{n'} (\pi/s)\,\, .
    \label{eq:TopologicalInvariant}
\end{equation}
In Fig.~\ref{fig:TopoInv}, we show the evolution of $\nu_n$ computed with Eq.~\eqref{eq:TopologicalInvariant} for the second and third bands, as the normalized modulation width $a/s$ grows. The first gap does not close for $a/s>0$, and therefore the first band is always in a topologically trivial phase. Simultaneously, we plot the size of the corresponding gap above each of these bands. As one would expect for a band topological invariant, the value of $\nu_n$ changes only when the gap above band $n$ closes, and the Zak phases of the crossing bands are exchanged. Since the invariant takes into account the sum of all Zak phases of the bands below, whenever two bands below gap $n$ cross and their Zak phases are exchanged, the topological invariant remains unchanged. For instance, when then second gap closes at the critical value $a_c\approx 0.44$, both $\gamma_2$ and $\gamma_3$ are exchanged, but their sum (mod $2\pi$) is still the same, such that $\nu_3$ is unaffected.

It is worth noting that the inclusion of small damping terms in the Drude conductivity of graphene, which introduces a small imaginary component in the frequencies of the modes (cf. Eq. \eqref{eq:ImaginaryDispersion}) does not seem to break the quantization of the Zak phase. Although the Hamiltonian would become non-Hermitian, the presence of inversion symmetry assures that the results for an Hermitian Zak phase are identical to the non-Hermitian Zak phase \cite{pocockTopologicalPlasmonicChain2018}

\subsection{\label{subsec:EdgeStates} Edge States }

\begin{figure}[t!]
\centering
\begin{minipage}[b]{0.45\textwidth}
 \includegraphics[width=\linewidth]{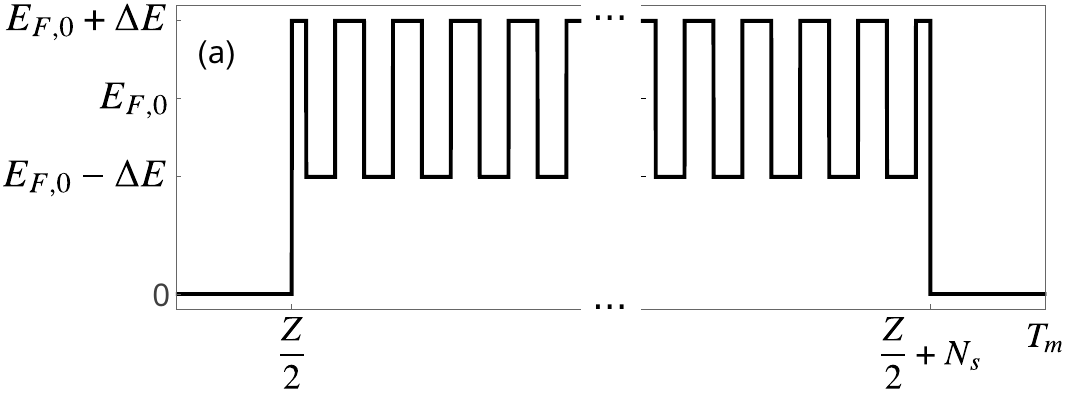}
\end{minipage}
\hfill
\vspace{0.8em}
\begin{minipage}[b]{0.48\textwidth}
    \centering
    \includegraphics[width=\linewidth]{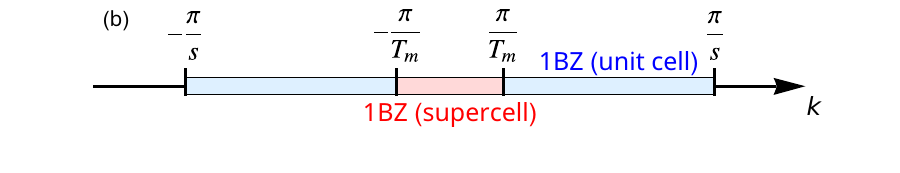}
\end{minipage}
    \caption{(a) Modulation profile of a single supercell. It has a region of plasmonic vacuum of length $Z$, where the Fermi energy is equal to zero, followed by a plasmonic crystal containing $N_s$ unit cells, equal to the ones used in the previous sections. (b) Sketch of the first BZ for the periodic lattice of the previous sections (in light blue) and the supercell (in light red). In the sketch, $T_m=5s$, but for large supercells the red interval becomes much smaller.}
    \label{fig:BZones}
\end{figure}

The bulk--boundary correspondence establishes a connection between a bulk topological invariant, such as $\nu_n$ defined above, and properties of the boundary of a finite-size system. In the present case of a 1D plasmonic crystal, this correspondence asserts the existence of mid-gap states localized at the system edges. The necessary condition to observe mid-gap states is an abrupt change of the topological index, which can be achieved at (a) a boundary of two plasmonic crystals in distinct topological phases, or (b) at a boundary between a crystal with non-trivial topological index and vacuum, which is always trivial.

To visualize the presence of these mid-gap states, we study the boundary of a plasmonic crystal with vacuum, by simulating a supercell containing $N_s$ unit cells together with a vaccum region of size $Z$ unit cells, where $E_{\mathrm{F}}$ is equal to zero~\footnote{By unit cell, it is meant the unit cell discussed in the previous sections, with length $s$, and not the supercell of length $T_m=(N_s+Z)s$}. The potential landscape $D(\boldsymbol{x})$ of a single supercell, which has a total size of $T_m=(Z+N_s)s$, is shown in Fig.~\ref{fig:BZones}a, with the finite crystal slab placed in the center of the supercell. In the numerical calculations, this supercell is periodically repeated, effectively forming an infinite array of slabs separated by vacuum regions.

\begin{figure}[t!]
\begin{minipage}[b]{0.45 \textwidth}
 \includegraphics[clip, trim=0.0cm 3.5cm 0cm 3.5cm ,width=\linewidth]{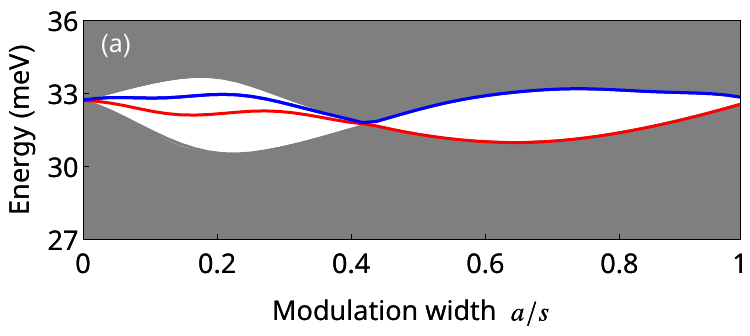}
\end{minipage}
\hfill
\begin{minipage}[b]{0.45\textwidth}
\raggedleft
 \includegraphics[clip, trim=0cm 0cm 0cm 0.5cm,width=\linewidth]{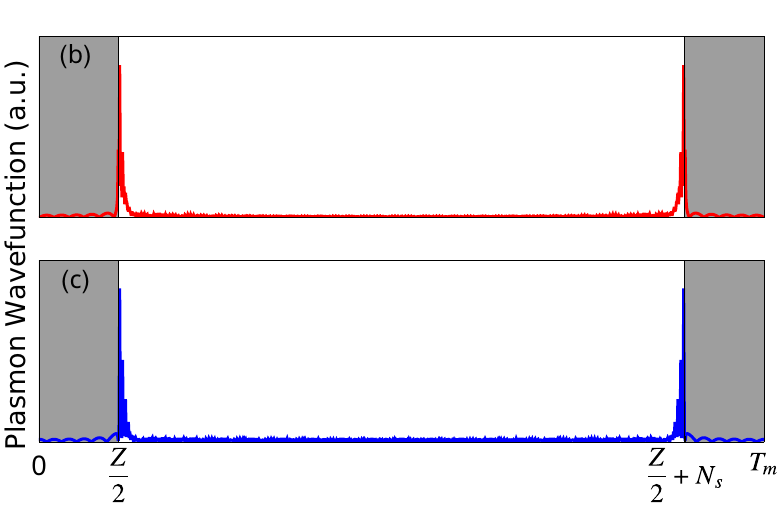}
\end{minipage}
\hfill
\caption{(a) Evolution of the spectrum in a finite plasmonic crystal with respect to the modulation width of a small unit cell. In gray are represented the bulk states of the second and third plasmonic bands. Highlighted in red and blue are the two edge states in the second gap. Below the topological phase transition, $\nu_2=\pi$ in the crystal bulk, and the finite crystal displays mid-gap states that merge into the bulk after the gap closes. (b) Wavefunction of the plasmon edge state highlighted in red at $a/s=0.2$. The plasmon wavefunction abruptly ends in the vacuum region (shaded in gray) and decays exponentially into the plasmonic crystal. (c) The plasmon edge state highlighted in blue. This state still shows some degree of localization at the edges, but interaction between the two ends of the crystal has shifted this state towards the bottom of the third band. Parameters used in the simulation were $Z = 56$ and $N_s=200$ (measured in units of $s$), with $2048$ total modes in the Hamiltonian}
\label{fig:EdgeState}
\end{figure}

The supercell modulation effectively changes the modulation profile, as in Eq.~\eqref{eq:DrudeModulation}, and since we still deal with an infinite periodic system, we may construct the Hamiltonian provided by Eqs.~\eqref{eq:FullHamiltonian}. However, the (Bloch) crystal momentum $\boldsymbol{k}$ now belongs to a mini-BZ on the interval $[-\pi/T_m,\pi/T_m]$ (c.f. Fig.~\ref{fig:BZones}b). 
Since we are considering the spectrum of a single crystal slab, this label becomes a numerical artifact that can be fixed to  $\boldsymbol{k}=0$. If each crystal is well isolated, then they host independent modes that become dispersionless with $\boldsymbol{k}$ as the period of the supercell $T_m$ increases. Thus, the Hamiltonian depends only on the reciprocal superlattice vectors $\boldsymbol{G}_n = 2\pi n /T_m$, which still label the plasmon crystal momentum and couple the plasmons through the Fourier coefficients of the supercell modulation $D_{\boldsymbol{G}}^{\text{SC}}$. The supercell Hamiltonian can be expressed in the following time-independent form
\begin{equation}
    \begin{split}
        H^{\text{SC}} = \frac{\hbar}{2}\sum_{\boldsymbol{G}} \biggl[\omega_{\boldsymbol{G}} a_{\boldsymbol{G}}a^\dagger_{\boldsymbol{G}}+& \biggr. \\
        \biggl.\sum_{\boldsymbol{G'}\neq\boldsymbol{G}}g^{\text{SC}}_{\boldsymbol{G},\boldsymbol{G'}}\parent{a_{\boldsymbol{G}}a_{\boldsymbol{G'}}^{\dagger}+a_{\boldsymbol{G}}a_{-\boldsymbol{G'}}}\biggr]+&  h.c.,
    \end{split}
    \label{eq:RealSpaceHamiltonian}
    \end{equation}
where we define the supercell coupling function as 
\begin{equation}
     g^{\text{SC}}_{\boldsymbol{G},\boldsymbol{G'}} =\frac{\hbar}{2}\frac{D^{\text{SC}}_{\boldsymbol{G}-\boldsymbol{G'}}}{D^{\text{SC}}_0}f(\boldsymbol{G})f(\boldsymbol{G'}).
    \label{eq:RealSpaceCouplingTerm}
\end{equation}
The function $f(\boldsymbol{q})$ is defined in Eq.~\eqref{eq:FormFactor}. In the limit $T_m\rightarrow \infty$, the reciprocal superlattice vectors become continuous and the Hamiltonian describes a truly isolated finite plasmonic crystal slab embedded in vacuum.

To establish a correspondence between the topology of the finite-size and infinite systems, the potential of the supercell at the edge of the crystal should be cut off at the same inversion symmetric point, which we have chosen to be the midpoint of the upper step of the modulation (cf. Figs.~\ref{fig:EFModulation} and \ref{fig:BZones}a). The vacuum length $Z$ is taken sufficiently large to suppress coupling between neighboring supercells, and the total number of modes figuring in the Hamiltonian is taken
large enough to resolve the crystal's unit cells.

In Fig.~\ref{fig:EdgeState}a, the evolution of the energy eigenvalues around the second gap is plotted as a function of the normalized modulation width $a/s$. As discussed in the previous sections, tuning this parameter induces a topological phase transition of the bulk topological invariant $\nu_2$. Since the vacuum is topologically trivial, mid-gap states should appear when the system is in the topological phase with $\nu_2=\pi$, which for the second band is below the gap closure (see Fig.~\ref{fig:TopoInv}). This is indeed the observed behavior of the states highlighted in red and blue in Fig.~\ref{fig:EdgeState}a. When the bulk states (gray shaded) are gapped, the mid-gap states lie within the gap, but as the two bands become gapless at the critical point and the topological invariant becomes trivial, the mid-gap states merge with the bulk of the two bands.

To further clarify that the mid-gap states are the edge-states predicted by the bulk--boundary correspondence, the amplitude of their wavefunctions, computed from Eq.~\eqref{eq:WavefunctionSpinor}, is also depicted in Figs.~\ref{fig:EdgeState}b-c for some $a<a_c$. Because we are considering only the positive-energy eigenvalues of the Hamiltonian, the wavefunction spinor is particle-like and has a vanishing hole component. The wavefunction also has negligible weight into the vaccum region of the supercell, shaded in gray. In contrast, the probability of finding the plasmon is maximized at the edge of the slab, and decays exponentially inside the crystal. This is the signature behavior of an edge state, which is also seen in other topological systems, most notably Su-Schrieffer-Heeger (SSH)-like systems~\cite{suSolitonsPolyacetylene1979,tatianag.rappoportTopologicalGraphenePlasmons2021,breyQuantumPlasmonsDouble2024,gorenTopologicalZakPhase2018}.

One should be careful to note that there should exist two degenerate edge states, one localized at each edge of the slab ~\cite{fuTopologicalInsulatorsInversion2007}. However, the relatively small size of the crystal slab used in the numerical diagonalization still allows for significant overlap between the two ends of the crystal, such that the edge states are simultaneously localized at both edges. Furthermore, the two edge states can, in this way, hybridize to form "bonding" and "anti-bonding" states, causing an energy splitting that is responsible for the energy difference between the two edge states.

\section{\label{sec:Conclusion} Conclusions}

In this article, we developed a formalism to study the quantum theory of plasmon excitations in structured media, applied to a system of stacked dielectric-graphene-metal that supports GSPs, when the optical conductivity of graphene is described by only a Drude-like term, and losses non-local effects are neglected. This setup is experimentally achievable, for instance by using an AuPd metal substrate \cite{lundebergTuningQuantumNonlocal2017}. We then showed that a periodic modulation of the Fermi energy level, which can be achieved by tuning the carrier density, couples plasmons of different momenta, leading to the formation of plasmonic bands. The dispersion of the quantized plasmons is shaped by the graphene--metal separation, which can lead to screened plasmons that have a linear dispersion $\omega_q\propto q$ when the separation is small, compared to the size of the unit cell.

Like in electronic systems, when the modulation is inversion symmetric, the geometric Zak (or Berry) phase of plasmon bands is quantized and can only take values of $0$ or $\pi$. We discussed how the value of this phase is connected to the parity eigenvalues of each band at the band center and edge. We further showed how the mechanism of band inversion occurs for the plasmonic bands, and defined a topological invariant that can distinguish different topological phases. In the 1D plasmonic crystal discussed, this is a $\mathbb{Z}_2$ invariant. Finally, we demonstrated how a non-trivial bulk topological phase is tied to the existence of mid-gap edge states via bulk-boundary correspondence. We numerically analyzed a finite-size slab of a plasmonic crystal in contact with vacuum, and observed mid-gap states whose wavefunction is exponentially localized at both the crystal's edges.

These results clarify how a periodic modulation of the properties of graphene can be used to tune topological band structures of GSPs. Finally, we underline that the theory presented in this paper lays down the foundations for quantum optics with  graphene plasmonic crystals exploring strongly confined screened plasmons.

\section{Acknowledgments}
NNRP acknowledges several discussions with Eduardo Castro on topological aspects of photonic crystals. This work was supported by the Portuguese Foundation for Science and Technology (FCT) in the framework of the Strategic Funding UID/04650/2025. A.O. also acknowledges financing through FCT PhD grant PD/BD/00681/2025. The Center for Polariton-driven Light--Matter Interactions (POLIMA) is sponsored by the Danish National Research Foundation (Project No. DNRF165).

\clearpage
\appendix
\onecolumngrid
\section{\label{appendix:QuantizationCalculations} Details on the quantization}

This appendix is dedicated to detailing the procedure of computing the total EM energy in a structured medium, which is behind \cref{eq:TotalEnergyFull,eq:CouplingOperator,eq:NonLinearEigenvalueEq,eq:TotalEnergySimplified} of the main text. Our approach follows the same spirit as that of Ref.~\cite{raymerQuantumTheoryLight2020}, but here we choose to quantize the vector potential modes instead of the 
fields $\boldsymbol{D}$ and $\boldsymbol{B}$.

\subsection{\label{subappendix:TotalEnergy} Total Energy}

As alluded to above Eq.~\eqref{eq:EnergyDensity}, the EM energy density has an electric $u_E$ and magnetic $u_B$ contribution, such that $u=u_E+u_B$. In frequency space, we can express these contributions in the following form. For the magnetic part,
\begin{align}
    u_{B}=\frac{1}{\mu_0}\int_{-\infty}^{t}dt'\,\,\boldsymbol{B}\cdot\frac{\partial\boldsymbol{B}}{\partial t} 
    =\frac{1}{2\mu_{0}}\int\int\frac{d\omega d\omega'}{2\pi}e^{-i(\omega-\omega')t}B^{\star}(\boldsymbol{r},\omega')B(\boldsymbol{r},\omega).
    \label{eq:MagneticEnergyDensity}
\end{align}
The electric contribution cannot be simplified in the same manner because the electric permittivity has added structure. However, it may still be expanded in the following form:
\begin{align}
    u_{E}&=\int_{-\infty}^{t}dt'\,\,\boldsymbol{E}(\boldsymbol{r},t)\cdot\partial_{t}\boldsymbol{D}(\boldsymbol{r},t) \nonumber\\
    &=\int\int\frac{d\omega d\omega'}{2\pi}E(\boldsymbol{r},\omega)\parent{-i\omega'}\epsilon_{0}\epsilon_{r}(\boldsymbol{r},\omega)E(\boldsymbol{r},\omega)\int_{-\infty}^{t}dt'\,\,e^{-i(\omega+\omega')t'}\nonumber\\
    &=\frac{1}{2}\int\int\frac{d\omega d\omega'}{2\pi}\left[\omega\epsilon_{0}\epsilon_{r}(\boldsymbol{r},\omega)+\omega'\epsilon_{0}\epsilon_{r}(\boldsymbol{r},\omega')\right]E(\boldsymbol{r},\omega)E(\boldsymbol{r},\omega')\parent{\frac{e^{-i(\omega+\omega')t}}{\omega+\omega'}}\nonumber\\
    &=\frac{1}{2}\int\int\frac{d\omega d\omega'}{2\pi}e^{-i(\omega-\omega')t}\parent{\frac{\omega\epsilon_{0}\epsilon_{r}(\boldsymbol{r},\omega)-\omega'\epsilon_{0}\epsilon_{r}(\boldsymbol{r},\omega')}{\omega-\omega'}}E^{\star}(\boldsymbol{r},\omega')E(\boldsymbol{r},\omega).
    \label{eq:ElectricEnergyDensity}
\end{align}
In going from the second to the third line, the integral over $t'$ is done by inserting a regulatory term $e^{-\delta t}$ and taking the limit $\delta \rightarrow 0$. In addition, the integrand is split into the sum of two identical terms, but with frequency labels swapped $\omega\leftrightarrow \omega'$. The last step is simply a change of variable $\omega'\rightarrow-\omega'$, and exploits that the electric field is real-valued, such that $E(\boldsymbol{r},-\omega)=E^{\star}(\boldsymbol{r},\omega)$ and $\epsilon_{r}(\boldsymbol{r},-\omega)=\epsilon_r(\boldsymbol{r},\omega)$.

To obtain an expression for the total EM energy $W$, the above expressions have to be further integrated over real space, yielding again two contributions $W=W_E+W_B$. To proceed, we plug \cref{eq:WeylGauge_E,eq:WeylGauge_B,eq:ModeDecompostion} into \cref{eq:MagneticEnergyDensity,eq:ElectricEnergyDensity}. We obtain the following expressions for the total magnetic and electric energies:
\begin{align}
        W_{B}&=\frac{1}{2\mu_{0}}\int\frac{d\boldsymbol{r}}{V}\sum_{q,q'}\alpha_{q'}\alpha_{q}\boldsymbol{A}_{q'}(\boldsymbol{r})\parent{\nabla\times\nabla\times{\boldsymbol{A}_{q}(\boldsymbol{r})}}e^{-i(\omega_{q}+\omega_{q'})t}+\alpha_{q'}\alpha_{q}^{\star}\boldsymbol{A}_{q'}(\boldsymbol{r})\parent{\nabla\times\nabla\times{\boldsymbol{A}_{q}^{\star}(\boldsymbol{r})}}e^{i(\omega_{q}-\omega_{q'})t}+c.c. \nonumber \\
        W_{E} &=\frac{\epsilon_{0}}{2}\int\frac{d\boldsymbol{r}}{V}\sum_{q,q'}-\alpha_{q'}\alpha_{q}\boldsymbol{A}_{q'}(\boldsymbol{r})\boldsymbol{A}_{q}(\boldsymbol{r})\omega_{q}\omega_{q'}\parent{\frac{\omega_{q}\epsilon_{r}(\boldsymbol{r},\omega_{q})+\omega_{q'}\epsilon_{r}(\boldsymbol{r},\omega_{q'})}{\omega_{q}+\omega_{q'}}}e^{-i(\omega_{q}+\omega_{q})t}\\
        &+\alpha_{q}^{\star}\alpha_{q'}\boldsymbol{A}_{q'}(\boldsymbol{r})\boldsymbol{A}_{q}^{\star}(\boldsymbol{r})\omega_{q}\omega_{q'}\parent{\frac{\omega_{q}\epsilon_{r}(\boldsymbol{r},\omega_{q})-\omega_{q'}\epsilon_{r}(\boldsymbol{r},\omega_{q'})}{\omega_{q}-\omega_{q'}}}e^{i(\omega_{q}-\omega_{q'})t}+c.c.
    \end{align}
We can rearrange these sums into two kinds of terms:
    \begin{align}
    \begin{split}
        W_{+}&=\frac{1}{2}\sum_{q,q'}\alpha_{q'}\alpha_{q}e^{-i(\omega_{q}+\omega_{q'})t}\int\frac{d\boldsymbol{r}}{V}\boldsymbol{A}_{q'}(\boldsymbol{r})\mc{D}^+\boldsymbol{A}_{q}(\boldsymbol{r})+c.c.\\
        W_{-}&=\frac{1}{2}\sum_{q,q'}\alpha_{q'}\alpha_{q}^{\star}e^{i(\omega_{q}-\omega_{q'})t}\int\frac{d\boldsymbol{r}}{V}\boldsymbol{A}_{q'}(\boldsymbol{r})\mc{D}^-\boldsymbol{A}_{q}^{\star}(\boldsymbol{r})+c.c.
        \end{split}
    \end{align}
These have the same form as presented in the main text in Eq.~\eqref{eq:TotalEnergyFull}, with the coupling operators $\mc{D}^\pm$ defined in Eq.~\eqref{eq:CouplingOperator}.

\subsection{\label{subappendix:OrthogonalityRelations} Orthogonality relations}

Using the generalized eigenvalue equation~\eqref{eq:NonLinearEigenvalueEq}, which is a direct application of Maxwell's curl equation of $\boldsymbol{H}$ in matter, one can construct what we call an orthogonality relation for the mode functions. For $\omega_q\pm\omega_{q'}\neq0$, we find
\begin{subequations}
\begin{equation}
    \int d\boldsymbol{r}\boldsymbol{A}_{q'}^{\star}\mc{D}^+\boldsymbol{A}_{q}=0
    \label{eq:OrthogonalityRelationsPlus}
\end{equation}
\begin{equation}
    \int d\boldsymbol{r}\boldsymbol{A}_{q'}D^-\boldsymbol{A}_{q}=0
    \label{eq:OrthogonalityRelationsMinus}
\end{equation}
\label{eq:OrthogonalityRelations}
\end{subequations}
The procedure to find both of these relations is identical, and the derivation similar to orthogonal weight functions of Sturm-Liouville theory~\cite{weberEssentialsMathMethods2013,raymerQuantumTheoryLight2020}. First, we rewrite the constitutive relation Eq.~\eqref{eq:NonLinearEigenvalueEq} as:
\begin{equation}
   \parent{\epsilon_{0}\omega_{q}^{2}\epsilon_{r}(\boldsymbol{r},\omega_{q})-\frac{1}{\mu_{0}}\curl \curl }\boldsymbol{A}_{q}(\boldsymbol{r})=0.
   \label{eq:RewrittenNonlinearEq}
\end{equation}
The recipe to obtain the relation in Eq.~\eqref{eq:OrthogonalityRelationsMinus} can be summarized in three steps: (1) multiply Eq.~\eqref{eq:RewrittenNonlinearEq} by $\omega_{q'}\boldsymbol{A}_{q'}(\boldsymbol{r})$ and integrate over space; (2) duplicate the resulting equation, take the complex conjugate and swap labels $q\leftrightarrow q'$ and integrate by parts twice; (3) Subtract the equations resultant from steps (1) and (2) and divide by $-(\omega_{q'}-\omega_q)$. This last step is not valid if $\omega_{q'}-\omega_q = 0$, which must be treated separately. To derive Eq.~\eqref{eq:OrthogonalityRelationsPlus}, the same procedure applies, except in step (2) one does not take the complex conjugate and in (3) the equations are summed and divided by $\omega_{q'}+\omega_q$, which does not suffer from the same problem if we assume $\omega_q > 0$.

Equation~\eqref{eq:OrthogonalityRelationsPlus} directly implies that $W_+ = 0$. To simplify the form of $W_-$, we assume that there are no accidental degeneracies, i.e. $\omega_q\neq\omega_{q'}$ for $q\neq q'$. The contribution of the remaining terms can be obtained by taking the limiting case $\omega_{q'}\rightarrow\omega_q$ of the coupling operator:

\begin{equation} \lim_{\omega_{q}\rightarrow\omega_{q'}}\rparent{\epsilon_{0}\omega_{q'}\omega_{q}\parent{\frac{\omega_{q}\epsilon_{r}(\boldsymbol{r},\omega_{q})-\omega_{q'}\epsilon_{r}(\boldsymbol{r},\omega_{q'})}{\omega_{q}-\omega_{q'}}}+\frac{1}{\mu_{0}}\nabla\times\nabla\times} \boldsymbol{A}_{q} = \epsilon_{0}\omega_{q}^2 \parent{\frac{\partial}{\partial\omega}\parent{\omega\epsilon_{r}(\boldsymbol{r},\omega)}\restriction_{\omega=\omega_{q}}+\epsilon_{r}(\boldsymbol{r},\omega_{q})}\boldsymbol{A}_{q},
\end{equation}
where we have used the constitutive relation Eq.~\eqref{eq:NonLinearEigenvalueEq} to rewrite the second term on the left side. Inserting this equation into the expression for the total energy yields the result presented in the main text, Eq.~\eqref{eq:TotalEnergySimplified}, which can be used to implement the canonical quantization of the mode amplitudes.

\subsection{Effective volume}

In the canonical quantization step of Eq. \ref{eq:CanonicalQuantizationStep}, it is necessary to introduce a normalizing constant, which we called an effective volume $V_\mathrm{eff}$, in order to reduce the Hamiltonian to that of a collection of simple quantum harmonic oscillators. Although this volume has no physical consequences, it can be thought of as the effective volume of a box where the electromagnetic field is being quantized. When the field operators $a_p$ and $a_p^\dagger$ are introduced into the expression for the classical energy, and this is compared to the Hamiltonian in Eq. \ref{eq:HarmonicOscilatorHamiltonian}, we have

\begin{equation}
\frac{1}{2}\sum_{p} \frac{\hbar \omega_p}{2V_{\mathrm{eff}}}a_{p}a_p^\dagger
    \int d\boldsymbol{r} \boldsymbol{A}_{p}(\boldsymbol{r}) \cdot \boldsymbol{A}_{p}^{\star}(\boldsymbol{r}) \left[\frac{\partial}{\partial\omega}\parent{\omega\epsilon_{r}(\boldsymbol{r},\omega)}+\epsilon_{r}(\boldsymbol{r},\omega_{p})
    \right]+c.c = \frac{1}{2}\sum_p \hbar\omega_q \parent{a_{p}a^\dagger_{p} + h.c.}.
\end{equation}

From the above equation, an expression for the normalizing length naturally follows

\begin{align}
    V_\mathrm{eff}&=\frac{1}{2} \int d\boldsymbol{r} \boldsymbol{A}_{p}(\boldsymbol{r}) \cdot \boldsymbol{A}_{p}^{\star}(\boldsymbol{r}) \left[\frac{\partial}{\partial\omega}\parent{\omega\epsilon_{r}(\boldsymbol{r},\omega)}+\epsilon_{r}(\boldsymbol{r},\omega_{p})
    \right]+c.c \nonumber\\
     &=\int d\boldsymbol{r}\boldsymbol{A}_{p} (\boldsymbol{r})\left[\epsilon_{r}(\boldsymbol{r},\omega_{p})+\frac{\omega_{p}}{2}\frac{\partial}{\partial\omega}\parent{\epsilon_{r}(\boldsymbol{r},\omega)}\right]\boldsymbol{A}_{p}^{\star}(\boldsymbol{r}).
     \label{eq:Appendix_Volume}
\end{align}

In the system of screened plasmons described in section \ref{subsec:AGSP}, because the system is assumed to homogeneous in the graphene plane, the dielectric function $\epsilon_r(\boldsymbol{r},\omega)$, and consequently also the mode functions $\boldsymbol{A}_p(\boldsymbol{r})$, depend only on the $z$ coordinate. Thus, the integrals over the $x,y$ coordinates in Eq. \eqref{eq:Appendix_Volume} will simply yield the area $S$ of the graphene sheet, where the vector potential is defined, which exactly cancels with the $1/\sqrt{S}$ prefactors introduced in eq. \eqref{eq:NewModeDecomposition}. Along the $z$ direction, the remaining integral, which now has dimensions of length, can be carried out using eqs.\eqref{eq:DielectricFunction} and \eqref{eq:AGSPModeFunctions}. The term in square brackets can be simplified to

\begin{equation}
    \frac{\partial}{\partial\omega}\parent{\omega\epsilon_{r}(\boldsymbol{r},\omega)}+\epsilon_{r}(\boldsymbol{r},\omega_{p}) = 2\epsilon_d,
\end{equation}

Substituting back into eq. \eqref{eq:Appendix_Volume}, with the integrations over the graphene plane already simplified, we obtain the effective length

\begin{align}
    L_{\mathrm{eff},q}&=2 \epsilon_d \parent{2\sinh{(qd)}^2\int_{0}^\infty e^{-2qz}\,dz \quad+\int_{-d}^{0} \sinh{\parent{q\parent{z+d}}}^2+\cosh{\parent{q\parent{z+d}}}^2 \,dz}\nonumber \\
    &=2\epsilon_d\parent{\frac{\sinh{(qd)}^2}{q}+\frac{1}{2q}\sinh{\parent{2qd}}}\nonumber \\
    &=\frac{2 \epsilon_d}{q}\parent{\sinh{\parent{qd}^2+\frac{1}{2}\sinh{\parent{2qd}}}},
\end{align}

which is presented as eq. \eqref{eq:AGSPEffLength} in the main text.

\subsection{\label{subappendix:PeriodicModulation} Periodic Modulation}

Let us fix the (generic) label $q$ to denote a 2D momentum label, and fix the polarization of the field to $p$-polarized modes, i.e. the electric field lies in the plane of incidence. Here we consider the effect of introducing a modulation in the dielectric function
in the expression for the total EM energy,
\begin{equation}
     \epsilon_{r}(\boldsymbol{r},\omega) = \epsilon_d +\sum_{\boldsymbol{G}}\epsilon_r^{\boldsymbol{G}}(z,\omega)e^{i\scalar Gx}.
\end{equation}
Because the magnetic contribution $W_B$ does not depend explicitly on the dielectric function, it is the same as the one derived before. However, the electric contribution will acquire an additional term $W_E\rightarrow W_E+W_{\text{int}}$ given by
\begin{equation}
    \begin{split}
        W_{int}&=\frac{\epsilon_{0}}{2}\sum_{\boldsymbol{q},\boldsymbol{G}\neq0}e^{i(\omega_{\boldsymbol{q}}-\omega_{\boldsymbol{q+G}})t'}\omega_{\boldsymbol{q+G}}\omega_{\boldsymbol{q}}\alpha_{\boldsymbol{q+G}}^{\star}\alpha_{\boldsymbol{q}}\int dzA_{\boldsymbol{q+G}}^{\star}(z)\mc R_{\boldsymbol{G}}^{-}(z,\boldsymbol{q},\boldsymbol{q+G})A_{\boldsymbol{q}}(z)\\&-e^{i(\omega_{\boldsymbol{q}}+\omega_{\boldsymbol{-(q+G)}})t'}\alpha_{\boldsymbol{-(q+G)}}\alpha_{\boldsymbol{q}}\omega_{\boldsymbol{-(q+G)}}\omega_{\boldsymbol{q}}\int dzA_{\boldsymbol{-(q+G)}}(z)\mc R_{\boldsymbol{G}}^{+}(z,\boldsymbol{q},-(\boldsymbol{q+G}))A_{\boldsymbol{q}}(z)+c.c.,
    \end{split}
\end{equation}
where we have defined a new operator
\begin{equation}
    \mc R_{\boldsymbol{G}}^{\pm}(z,\boldsymbol{q},\boldsymbol{q'})=\frac{\omega_{\boldsymbol{q}}\epsilon_{r}^{\boldsymbol{G}}(z,\omega_{\boldsymbol{q}})\pm\omega_{\boldsymbol{q'}}\epsilon_{r}^{\boldsymbol{G}}(z,\omega_{\boldsymbol{q'}})}{\omega_{\boldsymbol{q}}\pm\omega_{\boldsymbol{q'}}}.
    \label{eq:ModulatedCouplingOperator}
\end{equation}
For the case of a modulation of the Drude weight for graphene plasmons, this coupling operator takes the form
\begin{equation}
    \mc R^{\pm}_{\boldsymbol{G}}(z,\boldsymbol{q},\boldsymbol{q'})=\mp\frac{1}{\omega_{\boldsymbol{q'}}\omega_{\boldsymbol{q}}}\frac{D_{\boldsymbol{G}}}{\epsilon_{0}}\delta(z).
\end{equation}
When the mode functions are promoted to quantum-mechanical operators, this term becomes an interaction term given by
\begin{equation}
    H_{int}=\frac{\hbar}{2}\sum_{\boldsymbol{q},\boldsymbol{G}\neq0}\frac{D_{\boldsymbol{G}}A_{\boldsymbol{q}}(0)}{\epsilon_{0}\sqrt{\omega_{\boldsymbol{q}}\omega_{\boldsymbol{q+G}}L_{\boldsymbol{q}}L_{\boldsymbol{q+G}}}}\parent{e^{i(\omega_{\boldsymbol{q}}-\omega_{\boldsymbol{q+G}})t'}A_{\boldsymbol{q+G}}^{\star}(0)a_{\boldsymbol{q}}a_{\boldsymbol{q+G}}^{\dagger}+e^{i(\omega_{\boldsymbol{q}}+\omega_{\boldsymbol{-(q+G)}})t'}A_{\boldsymbol{q+G}}(0)a_{\boldsymbol{q}}a_{-(\boldsymbol{q+G})}}+c.c.,
\end{equation}
which has the same structure presented in the main text in Eqs.~\eqref{eq:FullHamiltonian}-\eqref{eq:FormFactor}.

\twocolumngrid
\bibliography{TopologicalPlasmons}
\end{document}